\documentclass[letterpaper, 10 pt, conference]{ieeeconf}  

\IEEEoverridecommandlockouts                              

\usepackage{amsmath}
\usepackage{amssymb}
\usepackage{xcolor}
\usepackage{comment}
\usepackage{tikz}
\usetikzlibrary{shapes,arrows}

\newcommand{\norm}[1]{\left\lVert#1\right\rVert}

\newtheorem{thm}{Theorem}

\newtheorem{lem}[thm]{Lemma}

\newtheorem{assum}[thm]{Assumption}

\newtheorem{defn}[thm]{Definition}

\newtheorem{rem}[thm]{Remark}

\title{\LARGE \bf
Robust constrained nonlinear Model Predictive Control with Gated Recurrent Unit model - Extended version}

\author{Irene Schimperna$^{1}$ and Lalo Magni$^{1}$
\thanks{$^{1}$Department of Civil and Architecture Engineering, University of Pavia, via Ferrata 3, Pavia, 27100, Italy
        {\tt\small irene.schimperna01@universitadipavia.it}, {\tt\small lalo.magni@unipv.it}}%
}

\begin{document}

\maketitle
\thispagestyle{empty}
\pagestyle{empty}

\begin{abstract}
In this paper we propose a robust Model Predictive Control where a Gated Recurrent Unit network model is used to learn the input-output dynamics of the system under control. Robust satisfaction of input and output constraints and recursive feasibility in presence of model uncertainties are achieved using a constraint tightening approach. Moreover, new terminal cost and terminal set are introduced in the Model Predictive Control formulation to guarantee Input-to-State Stability of the closed loop system with respect to the uncertainty term.

\textit{Keywords:} Nonlinear models; Control of constrained systems; Robust control of nonlinear systems; Optimal controller synthesis for systems with uncertainties; Neural networks technology.
\end{abstract}

\section{Introduction}
Model Predictive Control (MPC) \cite{Rawlings2019mpc-book} relies on accurate models to predict the dynamics of the system under control. 
In some situations obtaining a physical model of a nonlinear plant can be difficult and expensive, because the system under control is complex or because there is a lack of knowledge about the internal laws that govern the system. In this context it can be convenient to rely on black-box models, such as nonlinear ARX models \cite{DeNicolao1997NARX}, Koopman operator-based system identification \cite{Korda2018lkoopman-mpc}, Bayesian identification \cite{Piga2019performance-oriented-model} or Neural Network (NN) models \cite{Ren2022nn-modeling-mpc}. 
In recent years, the use of NN models in MPC has gained a growing popularity, thanks to the availability of large amounts of data and of the development of increasingly powerful techniques. In particular, when NN are used to describe the input-output relation of the plant, a class of networks that has shown remarkable performances are Recurrent Neural Networks (RNN) \cite{Bonassi2022rnn}. In fact this kind of networks have the same mathematical structure of a dynamical system, and are able to learn the temporal dependence between the input and the output of the plant. The two more successful RNN architectures are Long Short-Term Memory (LSTM) \cite{Hochreiter1997lstm} and Gated Recurrent Unit (GRU) \cite{Cho2014gru}. The performances of these two architectures are comparable and the choice of the best model depend on the application \cite{chung2014gru}, but GRU have a simpler structure and a lower number of trainable parameters for the same number of units.
For a comparison of RNN based MPC algorithms we refer to \cite{jung2023mpc-lstm}, where the practical aspects of the implementation of MPC using LSTM models are investigated, and different tools for the optimization are compared in terms of computation time. In \cite{Zarzycki2022mpc-gru-lstm} a computationally efficient MPC scheme that utilizes online advanced trajectory linearization is applied to LSTM and GRU models.

Another advantage of using LSTM or GRU models is that it is possible to constraint the training procedure to guarantee that the network fulfil some stability properties, such as Input-to-State Stability (ISS) and Incremental Input-to-State Stability ($\delta$ISS). In particular sufficient stability conditions have been derived in \cite{Terzi2021mpc-lstm} for LSTM and in \cite{Bonassi2021Stability-GRU} for GRU. Using networks satisfying such stability properties guarantees that, regardless of the initial condition of the internal state of the model, bounded inputs or disturbances lead to bounded states. This property is useful when designing a control scheme because it allows to design converging observers for the model states and can be exploited to design an MPC that guarantees closed loop stability.
On the basis of these properties, in \cite{Terzi2021mpc-lstm} an observer and a stabilizing MPC are proposed for LSTM models in presence of input constraints. The closed loop stability is proven in the nominal case, i.e. without considering model-plant mismatch.
In 
\cite{Bonassi2023mpc-deltaiss-gru} a closed-loop stable MPC  based on a $\delta$ISS GRU model is proposed. Firstly, an observer for the GRU model is developed, and then an MPC with input constraints is designed. Stability is proven under the assumption that the system behaves according to its GRU model, if the prediction horizon is larger than a given lower bound. However satisfying the requirement on the prediction horizon can be very conservative, leading to horizon lengths that are computationally intractable. 
In \cite{Bonassi2021offsetfree-mpc-gru} the authors propose an offset-free MPC based on a GRU model, that makes use of an integral term summed to the output of the MPC to guarantee tracking with null error. 

One of the reasons of the popularity of MPC is the possibility to explicitly include hard constraints in the control problem formulation. 
However, most of the NN based MPC formulations presented so far only allow to include input constraints, while there are limited theoretical results that includes state and output constraints. 
Indeed, in presence of model-plant mismatch or disturbances and state or output constraints, it is necessary to rely on robust MPC techniques to guarantee robust feasibility and constraint satisfaction despite the uncertainties. There exist many different approaches for robust MPC, both in a bounded deterministic framework \cite{Mayne2000constrained-mpc} and in a stochastic one \cite{Hewing2019cautious-mpc}. 
More specifically, it is possible to rely on tube MPC \cite{Langson2004Tube-MPC}, on constraint tightening methods \cite{Limon2002iss-mpc}, 
or on min-max approaches \cite{Raimondo2009minmax}. 
In \cite{Schimperna2023extended} a robust MPC based on LSTM models is proposed, where robustness is achieved using a constraint tightening technique. Moreover, the control scheme makes use of the offset-free design of \cite{Morari2012nonlinear-offset-free} to obtain tracking of time variant set points unknown in advance and zero error at steady state in presence of asymptotically constant set points and disturbances.
In \cite{Patan2018two-nn-mpc} the authors propose to first derive a Feed-Forward Neural Network (FFNN) model of the plant, and then use a second FFNN to estimate the uncertainty associated to the model. The error model is then used in the sythesis of a robust MPC.

In this paper we propose a robust MPC based on a GRU model to perform regulation at an equilibrium point while respecting input and output constraints. The schema also includes an observer, designed according to \cite{Bonassi2023mpc-deltaiss-gru}. All the model uncertainties are represented as an additive term on the output of the GRU model. Based on this representation, a constraint tightening is introduced in the MPC formulation to guarantee robust constraint satisfaction and recursive feasibility of the optimization problem. Moreover, a terminal cost and a terminal set are proposed on the basis of an incremental Lyapunov function of the GRU model. This formulation of the MPC allows to prove that the closed loop system is recursively feasible and ISS with respect to the uncertainty term. Notably, the proposed method does not require a large prediction horizon, allowing to keep the computational cost low. 

\subsection{Notation and stability definitions}
Considering a vector $v$, $v_{(j)}$ is its $j$-th component, $v^\top$ is its transpose, $\|v\|_2$ is its 2-norm, $\|v\|_\infty$ is its infinity-norm, $\|v\|^2_A = v^\top A v$ is the squared norm weighted with matrix $A$ and $\|v\|_A = \sqrt{v^\top A v}$. Inequalities between vectors are considered element by element. When $\|v\|$ is used it means that the expression is valid for any norm of vector $v$.
Given two vectors $v$ and $w$, $v \circ w$ is their element-wise product. 
Considering a matrix $M$, $M_{(ij)}$ is its element in position $ij$, $M_{(i*)}$ is its $i$-th row, $\lambda_{max}(M)$ and $\lambda_{min}(M)$ are its maximum and minimum eigenvalues, and $\|M\|_2$ and $\|M\|_{\infty}$ are its induced 2-norm and $\infty$-norm.
$\mathbf{0}_{m,n}$ is the $m \times n$ null matrix and $I_n$ is the $n \times n$ identity matrix.
For the definition of functions of classes $\mathcal{K}$, $\mathcal{K}_\infty$ and $\mathcal{KL}$ see \cite{Rawlings2019mpc-book}. 

The notions of ISS and $\delta$ISS are now introduced for the generic discrete-time dynamical system $x_{k+1} = f(x_k, u_k)$, with $x \in \mathbb{R}^{n_x}$ and $u \in \mathbb{R}^{n_u}$.
The stability notions are stated in the sets $\mathcal{X} \subseteq \mathbb{R}^{n_x}$ and $\mathcal{U} \subseteq \mathbb{R}^{n_u}$, with the set $\mathcal{X}$ assumed to be positive invariant, i.e. for any $u \in \mathcal{U}$, it holds that $x \in \mathcal{X} \implies f(x,u) \in \mathcal{X}$.

\begin{defn} [ISS]
    The system $x_{k+1} = f(x_k, u_k)$ is Input-to-State Stable (ISS) in the sets $\mathcal{X}$ and $\mathcal{U}$ if there exist functions $\beta \in \mathcal{KL}$ and $\gamma \in \mathcal{K}$ such that for any $k \geq 0$, any initial condition $x_0 \in \mathcal{X}$ and any input sequence $u_0, u_1, ..., u_{k-1}$ with $u_h \in \mathcal{U}$ for $h = 0, ..., k-1$, it holds that
    \begin{equation*}
        \| x_k \| \leq \beta(\|x_0\|, k) + \gamma \left( \max_{0 \leq h < k} \| u_h \| \right)
    \end{equation*}
\end{defn}

\begin{defn} [ISS-Lyapunov function]
    A continuous function $V: \mathcal{X} \to \mathbb{R}_{+}$ is an ISS-Lyapunov function for the system $x_{k+1} = f(x_k, u_k)$ if there exist functions $\alpha_1, \alpha_2, \alpha_3 \in \mathcal{K}_\infty$ and $\gamma \in \mathcal{K}$ such that for all $x \in \mathcal{X}$, $u \in \mathcal{U}$
    \begin{subequations} \label{eq:iss-lyap}
        \begin{align}
            & \alpha_1(\|x\|) \leq V(x) \leq \alpha_2(\|x\|) \label{eq:iss-lyap-pos-def} \\
            & V(f(x,u)) - V(x) \leq - \alpha_3 (\|x\|) + \gamma(\|u\|) \label{eq:iss-lyap-neg-def}
        \end{align}
    \end{subequations}
\end{defn}

\begin{lem} [\cite{Jiang2001iss}] \label{lem:iss-lyap-implies-iss}
   If the system $x_{k+1} = f(x_k, u_k)$ admits a continuous ISS-Lyapunov function in the sets $\mathcal{X}$ and $\mathcal{U}$, and $f(\cdot, \cdot)$ is continuous, then it is ISS in $\mathcal{X}$ and $\mathcal{U}$. 
\end{lem}

\begin{defn} [$\delta$ISS]
    The system $x_{k+1} = f(x_k, u_k)$ is Incrementally Input-to-State Stable ($\delta$ISS) in the sets $\mathcal{X}$ and $\mathcal{U}$ if there exist functions $\beta \in \mathcal{KL}$ and $\gamma \in \mathcal{K}_{\infty}$ such that for any $k \geq 0$, any pair of initial conditions $x_{a,0} \in \mathcal{X}$ and $x_{b,0} \in \mathcal{X}$, any pair of input sequences $u_{a,0}, u_{a,1}, ..., u_{a,k-1}$ and $u_{b,0}, u_{b,1}, ..., u_{b,k-1}$, with $u_{a,h}, u_{b,h} \in \mathcal{U}$ for all $h = 0,..,k-1$, it holds that
    \begin{equation*}
    \begin{split}
        \| x_{a,k} - x_{b,k} \| &\leq \beta(\|x_{a,0} - x_{b,0}\|, k) \\
        & + \gamma \left( \max_{0 \leq h < k} \| u_{a,h} - u_{b,h} \| \right)
    \end{split}
    \end{equation*}
\end{defn}

\section{Control algorithm}
\tikzstyle{block} = [draw, rectangle, 
    minimum height=3em, minimum width=5em]
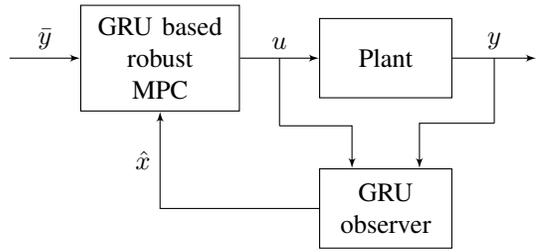
\begin{figure}
\centering
\begin{tikzpicture}[auto, node distance=2cm,>=latex']
    \node [coordinate] (input) {};
    \node [block, right of=input, align=center, minimum height=4em, minimum width=6em] (mpc) {GRU based\\robust\\MPC};
    \node [block, right of=mpc, node distance=3cm] (plant) {Plant};
    \node [block, below of=plant, align=center] (observer) {GRU\\observer};
    \node [coordinate, right of=plant] (output) {};

    \draw [->] (input) -- node {$\Bar{y}$} (mpc);
    \draw [->] (mpc) -- node [name=u] {$u$} (plant);
    \draw [->] (plant) -- node [name=y] {$y$} (output);
    \node [coordinate, below of=u, node distance=1.1cm] (u-below) {};
    \node [coordinate, below of=y, node distance=1.1cm] (y-below) {};
    \draw [-] (u) -- (u-below);
    \draw [-] (y) -- (y-below);
    \draw [->] (u-below) -| (observer.130);
    \draw [->] (y-below) -| (observer.50);
    \draw [->] (observer.180) -| node [pos=0.75] {$\hat{x}$} (mpc.270);
\end{tikzpicture}
\caption{Block diagram of the control scheme.} 
\label{fig:block-diagram}
\end{figure}

In this paper we propose a stabilizing MPC for a nonlinear plant, modelled by a GRU neural network. 
The objective of the control is to do regulation at a set point $\Bar{y}$, while respecting input saturation constraints 
\begin{equation}    \label{eq:input-constraint}
    u \in \mathcal{U} = \{ u \in \mathbb{R}^m : \| u \|_\infty \leq 1 \}
\end{equation}
and compact polytopic constraints on the output
\begin{equation}    \label{eq:output-constraint}
    y \in \mathcal{Y} = \{ y \in \mathbb{R}^p : Ly \leq h \}
\end{equation}
where $L \in \mathbb{R}^{q \times p}$ and $h \in \mathbb{R}^q$. 
Note that the formulation of the input saturation constraint as unity boundedness of $u$ is very general. In fact this formulation can be obtained by applying a proper normalization to the input of the model. This operation is a key element to speed up the training procedure and to obtain a well tuned NN.
The schematic layout of the proposed control algorithm is reported in Fig. \ref{fig:block-diagram}, and is composed by two main blocks.
The first block is an observer, that provides an estimation of the state of the GRU model on the basis of the input and the output of the plant. The second block is the robust MPC, that solves a Finite Horizon Optimal Control Problem (FHOCP) and gives in output the first element of the optimal input sequence. In the FHOCP the evolution of the plant is predicted by the GRU model, and the constraints are tightened to guarantee their robust satisfaction despite the presence of model-plant mismatch and of the observer estimation error. Moreover, the terminal ingredients of the FHOCP are tuned to guarantee stability.

\section{Gated Recurrent Unit model and observer} 

\subsection{Incrementally Input-to-State Stable Gated Recurrent Unit networks}
Gated Recurrent Units (GRU) neural networks are an architecture in the family of RNN. The GRU model has a state $x \in \mathbb{R}^n$, and, given the input $u \in \mathbb{R}^m$ of the plant, it produces an estimation $\xi \in \mathbb{R}^p$ of the output $y \in \mathbb{R}^p$ of the plant. The model equations are the following:
\begin{subequations}    \label{eq:gru}
    \begin{align}
        \begin{split}
          x_{k+1} &= z_k \circ x_k + (1 - z_k) \circ h_k 
        \end{split} \\
        z_k &= \sigma(W_z u_k + U_z x_k + b_z) \\
        r_k &= \sigma(W_r u_k + U_r x_k + b_r) \\
        h_k &= \tanh(W_h u_k + U_h (r_k \circ x_k) + b_h) \\
        \xi_k &= U_o x_k + b_o
    \end{align}
\end{subequations}
where $\sigma(x) = \frac{1}{1+e^{-x}}$ is the sigmoid activation function, applied element by element. Matrices $W_h, W_z, W_r \in \mathbb{R}^{n \times m}$, $U_h, U_z, U_r \in \mathbb{R}^{n \times n}$, $U_o \in \mathbb{R}^{p \times n}$ and vectors $b_h, b_z, b_r \in \mathbb{R}^n$, $b_o \in \mathbb{R}^p$ contain the trainable weights of the network. $z_k = z(x_k, u_k)$ is called update gate, while $r_k = r(x_k, u_k)$ is called forget gate or reset gate and $h_k = h(x_k, u_k)$ is the candidate activation.
The mathematical structure of the GRU model is the same of a discrete-time time-invariant dynamical system, that in a more compact way will be also denoted by
    \begin{align*}
        x_{k+1} &= f(x_k, u_k) \\
        \xi_k &= g(x_k)
    \end{align*}
In the following we recall a sufficient condition for the $\delta$ISS of the GRU model.
\begin{thm} [\cite{Bonassi2021Stability-GRU}] \label{th:stability-bonassi} 
    If the GRU model \eqref{eq:gru} is such that
    \begin{equation}    \label{eq:condition-deltaiss-gru}
        \nu < 1
    \end{equation}
    where
    \begin{equation*}
        \nu = \| U_h \|_\infty \left( \frac{1}{4} \| U_r \|_\infty + \Bar{\sigma}_r \right) + \frac{1}{4} \frac{1 + \Bar{\phi}_h}{1 - \Bar{\sigma}_z} \| U_z \|_\infty
    \end{equation*}
    and
        \begin{align*}
            \Bar{\sigma}_z &= \sigma(\norm{[W_z \quad U_z \quad b_z]}_\infty) \\
            \Bar{\sigma}_r &= \sigma(\norm{[W_r \quad U_r \quad b_r]}_\infty) \\
            \Bar{\phi}_h &= \tanh(\norm{[W_h \quad U_h \quad b_h]}_\infty)
        \end{align*}
    then it is $\delta$ISS in the sets $\mathcal{X} = \{ x \in \mathbb{R}^n : \| x \|_\infty \leq 1 \}$ and $\mathcal{U}$.
    \hfill $\square$
\end{thm}

Under the conditions of Theorem \ref{th:stability-bonassi}, it is possible to define an incremental Lyapunov function for the GRU model. This Lyapunov function is useful to define the terminal ingredients for the MPC and the parameters for the constraint tightening.

\begin{lem} \label{lem:Vs}
    If the GRU model respects condition \eqref{eq:condition-deltaiss-gru}, then
    \begin{equation*}
        V_s(x_a, x_b) = \norm{x_a - x_b}_\infty
    \end{equation*}
    is an incremental Lyapunov function for the system \eqref{eq:gru}, such that
    \begin{subequations} \label{eq:incremental-lyap}
        \begin{align}
            & V_s(x_a^+, x_b^+) \leq \rho_s V_s (x_a, x_b) \label{eq:incremental-lyap-neg-def} \\
            & L U_o (x_a - x_b) \leq c_{s} V_s (x_a, x_b)  \label{eq:incremental-lyap-constraints}
        \end{align}
    \end{subequations}
    where $x_a^+ = f(x_a, u)$, $x_b^+ = f(x_b, u)$, 
    $\rho_s = \bar{\sigma}_z + (1 - \bar{\sigma}_z) \nu$ and $c_s \in \mathbb{R}^q$ with $c_{s(j)} = \| (L U_o)_{(j*)} \|_\infty$.
\end{lem}

\textbf{Proof: } The proof is reported in the Appendix.

Note that if $\nu < 1$, then $\rho_s < 1$. In fact $\rho_s$ is the average between 1 and $\nu$ weighted by $\Bar{\sigma}_z \in (0,1)$. Hence, $\rho_s \in (\nu, 1)$.

The GRU model is not an exact representation of the dynamics of the system under control, because of the model-plant mismatch that is always present. We assume that the real plant with input $u$ and output $y$ behaves as a perturbed version of its GRU model, described by the following equations
\begin{subequations}    \label{eq:perturbed-model}
    \begin{align}
        x_{k+1} &= f(x_k, u_k) \label{eq:perturbed-model-state} \\
        y_k &= g(x_k) + w_{y,k} \label{eq:perturbed-model-output}
    \end{align}
\end{subequations}
where $w_y$ is bounded in the set
\begin{equation*}
    w_y \in \mathcal{W}_y = \{ w_y \in \mathbb{R}^p : \norm{w_y}_\infty \leq \Bar{w}_y \}
\end{equation*}
and $\Bar{w}_y$ can be estimated as the maximum output error of the model.

\subsection{Observer for the GRU model}
The state $x$ of the GRU model has no physical meaning, hence it cannot be directly measured from the system, but it needs to be estimated by means of an observer. In this work we use the observer for the GRU proposed in \cite{Bonassi2023mpc-deltaiss-gru}, that is described by the following equations:
\begin{subequations}    \label{eq:observer}
     \begin{align}
        \begin{split}
          \hat{x}_{k+1} &= \hat{z}_k \circ \hat{x}_k + (1 - \hat{z}_k) \circ \hat{h}_k
        \end{split} \\
        \hat{z}_k &= \sigma(W_z u_k + U_z \hat{x}_k + b_z + L_z (y_k - \hat{y}_k)) \\
        \hat{r}_k &= \sigma(W_r u_k + U_r \hat{x}_k + b_r + L_r (y_k - \hat{y}_k)) \\
        \hat{h}_k &= \tanh(W_h u_k + U_h (\hat{r}_k \circ \hat{x}_k) + b_h) \\
        \hat{y}_k &= U_o \hat{x}_k + b_o
    \end{align}
\end{subequations}
where $L_z, L_r \in \mathbb{R}^{n \times p}$ are observer gains that must be properly selected, $\hat{z}_k = \hat{z}(\hat{x}_k, u_k, y_k)$, $\hat{r}_k = \hat{r}(\hat{x}_k, u_k, y_k)$ and $\hat{h}_k = \hat{h}(\hat{x}_k, u_k, y_k)$.
In presence of perturbation it is not possible to guarantee that the observer estimation error converges to zero. However it is possible to prove that it is ISS with respect to the perturbation term $w_y$.

\begin{lem} \label{lem:Vo}
    If the plant behaves according to \eqref{eq:perturbed-model} and respects the $\delta$ISS condition \eqref{eq:condition-deltaiss-gru}, $x \in \mathcal{X}$, $\hat{x} \in \mathcal{X}$, $u \in \mathcal{U}$ and the observer gains $L_z$ and $L_r$ are selected so that
    \begin{equation}    \label{eq:condition-observer-convergence}
        \nu_o < 1
    \end{equation}
    where
    \begin{equation*}
        \begin{split}
            \nu_o &= \| U_h \|_\infty \left( \frac{1}{4} \| U_r - L_r U_o \|_\infty + \Bar{\sigma}_r \right) \\
            &+ \frac{1}{4} \frac{1 + \Bar{\phi}_h}{1 - \Bar{\sigma}_z} \| U_z - L_z U_o \|_\infty
        \end{split}
    \end{equation*}
    then the function
    \begin{equation*}
        V_o (x - \hat{x}) = \| x - \hat{x} \|_\infty
    \end{equation*}
    is an ISS-Lyapunov function for the system describing the observer estimation error $\hat{x} - x$ with respect to the input $w_y$, such that
    \begin{subequations}    \label{eq:lyap-observer-properties}
        \begin{align}
            &V_{o}(x^+ - \hat{x}^+) \leq \rho_o V_{o}(x - \hat{x}) + \kappa \norm{w_y}_\infty \label{eq:lyap-observer-neg-def} \\
            &LU_o (x - \hat{x}) \leq c_o V_{o}(x - \hat{x})  \label{eq:lyap-observer-constraint} \\
            &\norm{\hat{x}^+ - f(\hat{x}, u)}_\infty \leq L_{max} V_{o}(x - \hat{x}) + \kappa \norm{w_y}_\infty \label{eq:lyap-observer-max-gain}
        \end{align}
    \end{subequations}
    where 
    $x^+ = f(x,u)$, $\hat{x}^+$ is the next state computed by the observer \eqref{eq:observer},
    $\rho_o = \Bar{\sigma}_z + (1 - \Bar{\sigma}_z) \nu_o$, 
    $c_o \in \mathbb{R}^q$ with $c_{o(j)} = \| (L U_o)_{(j*)} \|_\infty$, and
    \begin{equation*}
        \kappa = \frac{1}{4} (1 + \Bar{\phi}_h) \| L_z \|_\infty  + \frac{1}{4} \Bar{\sigma}_z \| U_h \|_\infty \| L_r \|_\infty
    \end{equation*}
    \begin{equation*}
        L_{max} =  \frac{1}{4} (1 + \Bar{\phi}_h) \norm{L_z U_o}_\infty + \frac{1}{4} \Bar{\sigma}_z \norm{U_h}_\infty \norm{L_r U_o}_\infty
    \end{equation*}
\end{lem}

\textbf{Proof: } The proof is reported in the Appendix.

\begin{rem} \label{rem:observer-gains}
    A possible suboptimal choice of the gains of the observer that satisfies the condition $\nu_o < 1$ is the open loop observer with $L_z = L_r = \mathbf{0}_{n,p}$. In fact with this choice $\nu_o = \nu$, that is smaller than 1 by assumption.
\end{rem}

\begin{thm} \label{th:iss-observer}
    If the plant behaves according to \eqref{eq:perturbed-model} and respects the $\delta$ISS condition \eqref{eq:condition-deltaiss-gru}, $x \in \mathcal{X}$, $\hat{x} \in \mathcal{X}$, $u \in \mathcal{U}$, $w_y \in \mathcal{W}_y$ and $\nu_o < 1$, then the system that describes the observer estimation error $x - \hat{x}$ is ISS with respect to the input $w_y$.
\end{thm}

\textbf{Proof: } The proof follows from Lemma \ref{lem:iss-lyap-implies-iss}, in view of the existence of an ISS-Lyapunov function $V_o(x - \hat{x})$.

\section{MPC design}
\subsection{Tightened constraint design}

On the basis of the derived Lyapunov functions for the model and for the observer estimation error, it is possible to introduce the coefficients $a_i, b_i \in \mathbb{R}^{q}$ for the constraint tightening that will be employed in the robust MPC. Such coefficients are designed as follows \cite{Kohler2019simple-robust-MPC}:
\begin{subequations}    \label{eq:constraint_tightening_parameters}
    \begin{align}
        a_{0} & = c_{o}, \quad b_0 = \mathbf{0}_{q,1} \label{eq:initialization_a_ji} \\
        a_{i+1} & = \rho_o a_{i} +  \rho_s^i L_{max} c_{s} \label{eq:recursion_a_ji} \\
        b_{i+1} &= b_i + a_i \Bar{w} + c_s \rho_s^i \Bar{w}    \label{eq:recursion_b_i}
    \end{align}
\end{subequations}
where $\Bar{w}$ is such that
\begin{equation}    \label{eq:w-bar}
    \kappa \norm{w_y}_\infty \leq \Bar{w}
\end{equation}
for all $w_y \in \mathcal{W}_y$ .

For the constraint tightening the MPC uses also a time variant term $\hat{e}_o \in \mathbb{R}$ related to the uncertainty of the observer. 
The evolution in time of this term corresponds to the worst case evolution of the observer ISS-Lyapunov function $V_o(x - \hat{x})$:
\begin{equation}    \label{eq:eo-evolution}
     \hat{e}_{o,k+1} = \rho_o \hat{e}_{o,k} + \Bar{w}
\end{equation}
Note that depending on the values of $\hat{e}_{o,0}$, $\rho_o$ and $\Bar{w}$, $\hat{e}_{o}$ can increase or decrease, but its behavior is always monotonic with
\begin{equation*}
    \lim_{k \to \infty} \hat{e}_{o,k} = \Bar{e}_\infty = \frac{\Bar{w}}{1 - \rho_o}
\end{equation*}

\subsection{Robust MPC formulation}
It is now introduced the Finite Horizon Optimal Control Problem (FHOCP) solved by the MPC. In the FHOCP it is penalized the deviation of states and inputs from the reference values $\Bar{x}$ and $\Bar{u}$. These references values are computed from the GRU model \eqref{eq:gru} as the equilibrium state and input corresponding to the output $\xi = \Bar{y}$.

\begin{defn} [FHOCP]
    Given the prediction horizon $N$, the FHOCP for the robust MPC is the following:
    \begin{subequations} \label{eq:optimization}
        \begin{align}
            \begin{split}
                \min_{u_{\cdot|k}} & \sum_{i=0}^{N-1} \left( \norm{x_{i|k} - \Bar{x}}_Q^2 + \norm{u_{i|k} - \Bar{u}}_R^2 \right) \\
                &+ s \norm{x_{N|k} + \Bar{x}}_\infty^2
            \end{split}  \label{eq:optimization-cost} \\
            \textrm{s.t.} \quad & x_{0|k} = \hat{x}_k \label{eq:optimization-initialization} \\
            & x_{i+1|k} = f(x_{i|k}, u_{i|k}) \\
            & L(U_o x_{i|k} + b_o) + w_L \leq h - a_{i} \hat{e}_{o,k} - b_i  \label{eq:optimization-constraint} \\
            & u_{i|k} \in \mathcal{U} \label{eq:optimization-input-constraint} \\
            & \text{for } i = 0, ..., N-1 \\
            & x_{N|k} \in \mathcal{X}_f \label{eq:optimization-terminal-constraint} 
        \end{align}
    \end{subequations}
    where $Q$ and $R$ are positive definite matrices and are design choices, while
    \begin{equation}    \label{eq:definition_s}
        s \geq \frac{n \lambda_{max}(Q)}{1 - \rho_s^2}
    \end{equation}
    The term $w_L \in \mathbb{R}^q$ for the constraint tightening is defined element by element as
    \begin{equation}
        w_{L(j)} = \sum_{k=1}^p |L_{(jk)}| \Bar{w}_y
    \end{equation}
    for $j = 1, ..., q$.
    
    $\mathcal{X}_f$ is a terminal set chosen as a level line of the terminal cost:
    \begin{equation}    \label{eq:terminal_constraint}
        \mathcal{X}_f = \left\{ x \in \mathbb{R}^n : \norm{x - \Bar{x}}_\infty \leq \alpha \right\}
    \end{equation}
    with
    \begin{subequations}    \label{eq:def-alpha}
        \begin{equation}
            \alpha = \min_{j = 1,...,q} \frac{-L_{(j*)} \Bar{y} + h_{(j)} - \tilde{e}_{o} a_{N(j)} - b_{N(j)} - w_{L(j)}}{\norm{(L U_o)_{(j*)}}_\infty}
        \end{equation}
        where 
        \begin{equation}
            \tilde{e}_{o} = \max \{ \hat{e}_{o,0}, \Bar{e}_\infty \}
        \end{equation}
    \end{subequations}
    \hfill $\square$
\end{defn}

At each time step $k$ the solution of the FHOCP is denoted by $u^*_{0|k},...,u^*_{N-1|k}$. According to the Receding Horizon principle, the MPC control law is obtained applying only the first element of the optimal input sequence:
\begin{equation}    \label{eq:mpc}
    u_k = k^{MPC}(\hat{x}_k, \hat{e}_{o,k}) = u^*_{0|k}
\end{equation}

\subsection{Recursive feasibility and stability analysis}
In this Subsection it is analysed the recursive feasibility and stability of the proposed control schema.
First note that in order to have a solution of the FHOCP, it is necessary that $\alpha > 0$. To guarantee this condition we introduce the following assumption on the set-point.
\begin{assum}   \label{ass:setpoint}
    The set-point $\Bar{y}$ is such that
    \begin{equation*}
        L \Bar{y} < h - \Tilde{e}_{o} a_{N} - b_{N} - w_L
    \end{equation*}
    \hfill$\square$
\end{assum}
Let's define the state of the closed loop system, shifted with respect to the nominal equilibrium, as $\psi = [(x - \Bar{x})^\top \; (\hat{x} - \Bar{x})^\top \; (\hat{e}_o - \Bar{e}_\infty)]^\top$
and the feasible set of states
\begin{equation}    \label{eq:X-mpc}
    \begin{split}
        \mathcal{X}^{MPC} = & \{ \psi : x, \hat{x} \in \mathcal{X}, \hat{e}_o \text{ is such that } V_o(x - \hat{x}) \leq \hat{e}_o \\
        & \text{and } \exists \text{ a solution of the FHOCP} \}
    \end{split}   
\end{equation}
Then it is possible to state the following theorem concerning the properties of the closed loop system.
\begin{thm} \label{th:feasibility-stability}
    If the plant behaves according to \eqref{eq:perturbed-model}, the GRU model respects the $\delta$ISS condition \eqref{eq:condition-deltaiss-gru}, the observer gains are selected to satisfy condition \eqref{eq:condition-observer-convergence}, Assumption \ref{ass:setpoint} is satisfied and the following inequality holds
    \begin{equation}    \label{eq:condition-recursive-feasibility}
        \rho_s^N \left( L_{max} \hat{e}_{o,k} + \Bar{w} \right) \leq \alpha (1 - \rho_s)
    \end{equation}
    for all $k\geq 0$, then the FHOCP is recursively feasible,
    the constraints \eqref{eq:input-constraint} and \eqref{eq:output-constraint} are satisfied $\forall \psi \in \mathcal{X}^{MPC}$ and the closed loop system is ISS with respect to the input $w_y$, for $\psi \in \mathcal{X}^{MPC}$ and $w_y \in \mathcal{W}_y$.
\end{thm}

\textbf{Proof: } The proof is reported in the Appendix.

\begin{rem}
    The satisfaction of inequality \eqref{eq:condition-recursive-feasibility} depends on the values of $N$, $L_{max}$ and $\Bar{w}$. The value of $L_{max}$ is related to the observer gains $L_z$ and $L_r$, that can be selected as small as needed, as noted in Remark \ref{rem:observer-gains}. The term $N$ can be enlarged to satisfy the condition for recursive feasibility. However with large values of $N$ the requirement of Assumption \ref{ass:setpoint} may become more conservative, and the computation effort needed to solve the optimization increases. Lastly, $\Bar{w}$ is related to the maximum of the perturbation term $w_y$, that represent the modeling error. Hence recursive feasibility can be obtained for any value of the prediction horizon $N$ when the modeling error $w_y$ is small enough, by using a proper tuning of the observer gains.
\end{rem}

\section{Numerical example}
\begin{figure}
\begin{center}
\includegraphics[width=7.5cm]{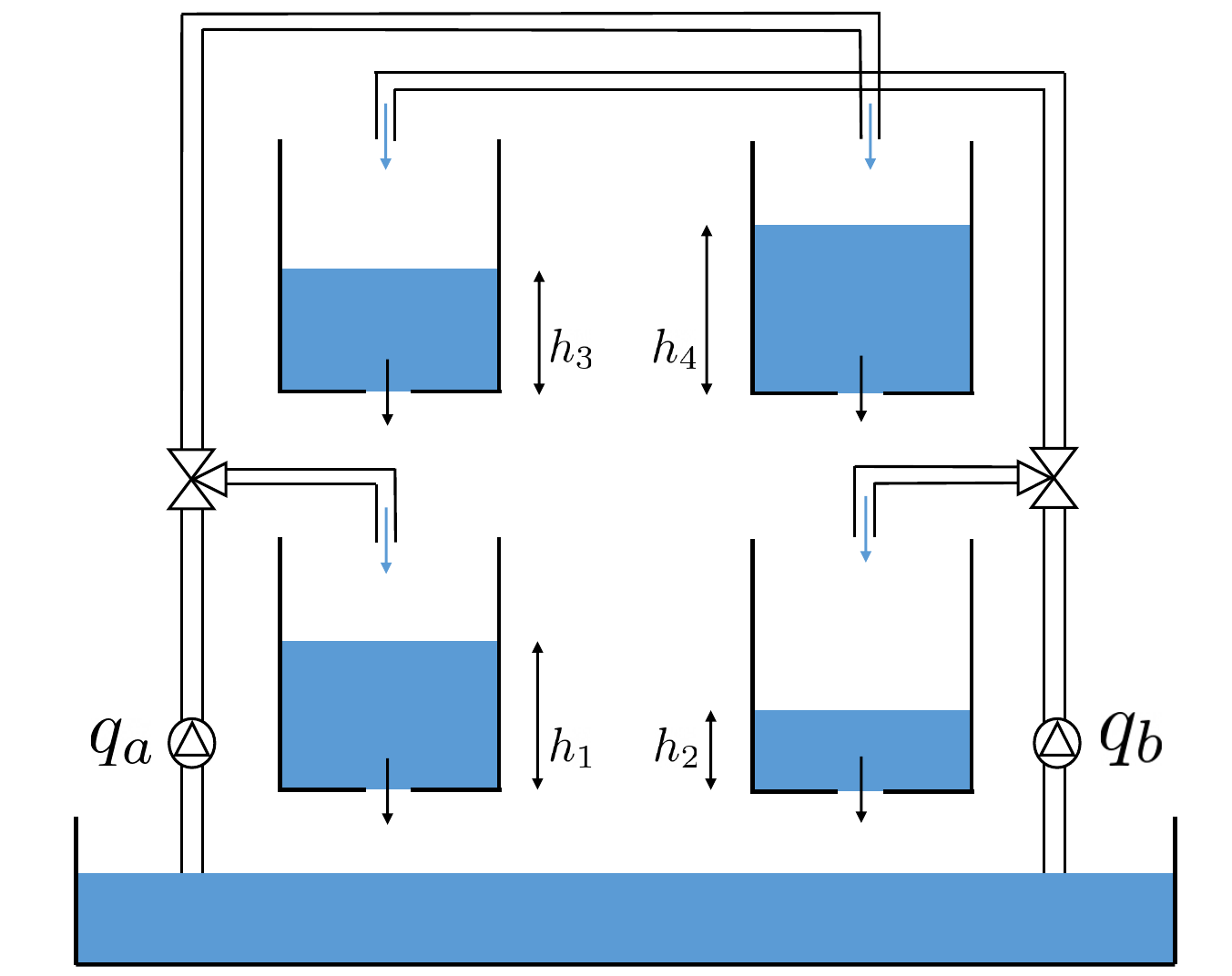}    
\caption{Schematic layout of the four tanks plant.}  
\label{fig:four-tanks}                                 
\end{center}                                 
\end{figure}

As numerical example to test the proposed control algorithm we consider the four tank benchmark described in \cite{alvarado2011four-tank}. The schematic layout of the plant is reported in Fig. \ref{fig:four-tanks}. The system is characterized by the following differential equations:
    \begin{align*}
        \Dot{h}_1 &= - \frac{a_1}{S} \sqrt{2gh_1} + \frac{a_3}{S} \sqrt{2gh_3} + \frac{\gamma_a}{S} q_a \\
        \Dot{h}_2 &= - \frac{a_2}{S} \sqrt{2gh_2} + \frac{a_4}{S} \sqrt{2gh_4} + \frac{\gamma_b}{S} q_b \\
        \Dot{h}_3 &= - \frac{a_3}{S} \sqrt{2gh_3} + \frac{1 - \gamma_b}{S} q_b \\
        \Dot{h}_4 &= - \frac{a_4}{S} \sqrt{2gh_4} + \frac{1 - \gamma_a}{S} q_a 
    \end{align*}
where the states $h_1, h_2, h_3, h_4$ are the water level in the four different tanks, while the inputs are $q_a$ and $q_b$ and represent the inlet flows in the two valves. The numerical values of the system parameters can be found in \cite{alvarado2011four-tank}. The objective is to control the water level in the two bottom tanks, $h_1$ and $h_2$, while respecting the following constraints:
\begin{align*}
    &q_a \in [0.0, 9.05]\times10^{-4} m^3/s &h_1 \in [0.0, 2.0] m  \\
    &q_b \in [0.0, 11.1]\times10^{-4} m^3/s &h_2 \in [0.0, 2.0] m
\end{align*}

To obtain the dataset for the training of the GRU network a simulator of the plant was forced with a multilevel pseudo-random signal, and the output response was sampled with a sampling time $T_s = 25s$.
In particular $q_a$ and $q_b$ in the dataset are two piecewise constant signals, with random values within the saturation constraints and a random changing step time.
A sequence of input-output data of 20000 time steps was collected, and it was split in 15000 time steps for training, 2500 for validation and 2500 for testing. Then, the training data have been divided into shortest subsequences of length 500 time steps each, and the training was performed using a batch size of 5. To enforce the $\delta$ISS property in the final network, a regularization terms was introduced in the loss to penalize sets of weights that do not respect the condition $\nu < 1$, as suggested in \cite{Bonassi2022rnn}. The performances of the final GRU model have been assessed with the FIT index, defined as
\begin{equation*}
    FIT = 100 \left( 1 - \frac{\norm{y - \xi}_2}{\norm{y - y_{avg}}_2} \right) \%
\end{equation*}
where $y$ is the real output present in the dataset, $\xi$ is the output predicted by the model and $y_{avg}$ is the average value of $y$.
The final network has $n=20$ neurons, respects the $\delta$ISS condition and has a FIT on the test dataset of 92.7\% for $h_1$ and of 89.7\% for $h_2$. 
For the simulation the observer gains $L_r$ and $L_z$ have been selected to minimize $\nu_o$, that is equivalent to minimize $\norm{U_r - L_r U_o}_\infty$ and $\norm{U_z - L_z U_o}_\infty$. The resulting gains are such that $\norm{L_r}_\infty = 0.002$ and $\norm{L_z}_\infty = 0.003$.
The cost matrices for the MPC were set to $Q = I_n$ and $R = 0.01 I_m$, and it was selected a prediction horizon of $N = 15$. $\Bar{w}_y$ was estimated as the maximum absolute value of the prediction error on the test dataset in the normalized variables, obtaining $\Bar{w}_y = 0.12$.
To try to keep as low as possible the initial observer estimation error, the initial output of the plant was measured and $\hat{x}_0$ was set to the equilibrium point of the GRU network corresponding to that output. The initial value of $\hat{e}_o$ was assumed to be 0.02, and it monotonically decreases to $\Bar{e}_\infty \approx 0.01$. 
\begin{figure}
\begin{center}
\includegraphics[width=8.4cm]{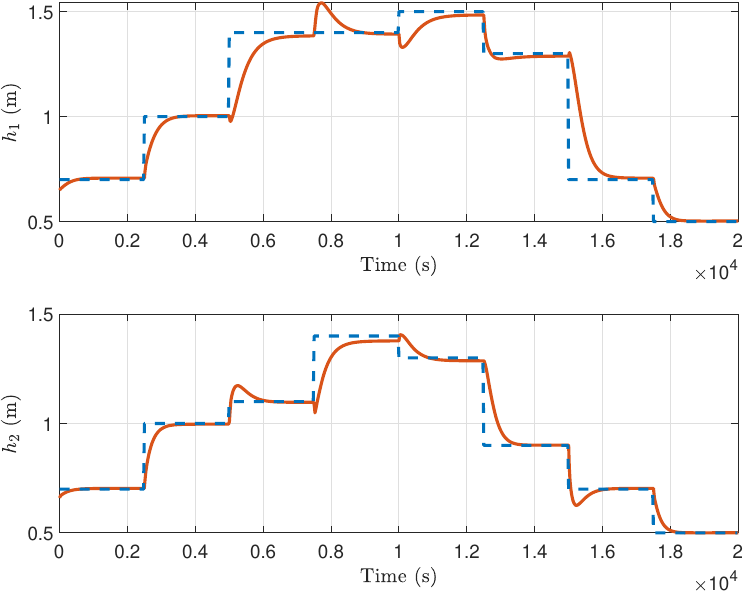}    
\caption{Closed loop performances: output (solid red line) and references (dashed blue line).}  
\label{fig:output}                                 
\end{center}                                 
\end{figure}
The simulation was performed by applying step-wise variations of the reference $\Bar{y}$, where all the values of the reference respect Assumption \ref{ass:setpoint}. When $\Bar{y}$ changes, the value of $\alpha$ that defines the terminal set is updated. The simulation parameters are such that condition \eqref{eq:condition-recursive-feasibility} is satisfied for all $k \geq 0$. 
The FHOCP was solved with \textit{fmincon} function of Matlab optimization toolbox.
The resulting closed loop trajectories are reported in Fig. \ref{fig:output}, showing that the controller is able to manage the plant in a robust satisfactory way in spite of uncertainties due to modelling errors. Only a small static gain mismatch is present.

\section{Conclusion}
This paper develops and analyses a robust MPC with input and output constraints, based on the use of a GRU model. Robustness and recursive feasibility are achieved using a constraint tightening approach, based on Lyapunov functions for the model and for the observer estimation error. The MPC formulation includes terminal ingredients that allow to guarantee ISS of the closed loop system with respect to the perturbation terms without having requirements on the length of the prediction horizon, allowing to maintain the computational effort tractable. 
In the paper, the algorithm has been proposed for regulation around an equilibrium point, but can be generalized to tracking, penalising also the control change in the MPC cost function.
Zero error reference tracking can be achieved by following the guidelines for nonlinear offset-free MPC of \cite{Morari2012nonlinear-offset-free}. This approach was used in a robust MPC based on LSTM models in \cite{Schimperna2023extended}, and the extension to the GRU based MPC proposed in this work is straightforward.


\bibliographystyle{unsrt}        
\bibliography{bibliography}


\appendix
\section{Proofs}    
For sake of readability, in some of the proofs the time index $k$ will be omitted, and the quantities at the subsequent time step will be denoted with the superscript $^+$.

The following property will be used in the proofs.
\begin{lem}
    Given two vectors $x, y \in \mathbb{R}^n$
    \begin{equation}    \label{eq:square-infty-norm}
        \norm{x + y}_\infty^2 \leq \norm{x}_\infty^2 + \norm{y}_\infty^2 + 2 \norm{x \circ y}_\infty
    \end{equation}
\end{lem}
\textbf{Proof:}
\begin{equation*}
    \begin{split}
        & \norm{x + y}_\infty^2 = \left( \max_{i} |x_{(i)} + y_{(i)}| \right)^2 = \max_{i} (x_{(i)} + y_{(i)})^2 \\
        & = \max_{i} (x_{(i)}^2 + y_{(i)}^2 + 2 x_{(i)} y_{(i)}) \\
        & = \max_{i} |x_{(i)}^2 + y_{(i)}^2 + 2 x_{(i)} y_{(i)}| \\
        & \leq \max_{i} |x_{(i)}^2| + \max_{i} |y_{(i)}^2| + 2 \max_{i} |x_{(i)} y_{(i)}|\\
        & = \norm{x}_\infty^2 + \norm{y}_\infty^2 + 2 \norm{x \circ y}_\infty
    \end{split}
\end{equation*}
\hfill $\square$

\subsection{Proof of Lemma \ref{lem:Vs}}
Preliminarily note that if $x \in \mathcal{X}$ and $u \in \mathcal{U}$, then $\Bar{\sigma}_z$, $\Bar{\sigma}_r$ and $\Bar{\phi}_h$ are bounds for $z$, $r$ and $h$, as shown in \cite{Bonassi2021Stability-GRU}. In particular we have that
\begin{subequations}    \label{eq:gates-bounds}
    \begin{align}
        |z_{(j)}| &\leq \norm{z}_\infty \leq \Bar{\sigma}_z \\
        |r_{(j)}| & \leq \norm{r}_\infty \leq \Bar{\sigma}_r \\
        |h_{(j)}| & \leq \norm{h}_\infty \leq \Bar{\phi}_h
    \end{align}
\end{subequations}
Moreover, in view of the symmetry of the sigmoid activation function $\sigma(\cdot)$, we have that for $j = 1,...,n$
\begin{subequations} \label{eq:bound-z}
    \begin{align}
        & 0 < 1 - \Bar{\sigma}_z \leq z_{(j)} \leq \Bar{\sigma}_z < 1 \label{eq:bound-z-1} \\
        & 0 < 1 - \Bar{\sigma}_z \leq 1 - z_{(j)} \leq \Bar{\sigma}_z < 1 \label{eq:bound-z-2}
    \end{align}
\end{subequations}

\textit{Condition \eqref{eq:incremental-lyap-neg-def}:}
Let's start by denoting $z_a = z(x_a, u)$, $r_a = r(x_a, u)$, $h_a = h(x_a, u)$, and let's adopt the same notation for $z_b, r_b, h_b$. Moreover, let $\Delta x = x_a - x_b$. Let's now consider the time evolution of the $j$-th component of $\Delta x$:
\begin{equation*}
    \begin{split}
        &\Delta x^+_{(j)} = z_{a(j)} x_{a(j)} + (1 - z_{a(j)}) h_{a(j)} \\
        & - z_{b(j)} x_{b(j)} - (1 - z_{b(j)}) h_{b(j)} \\
        & = z_{a(j)} (x_{a(j)} - x_{b(j)}) + (z_{a(j)} - z_{b(j)}) x_{b(j)} \\
        & + (1 - z_{a(j)}) (h_{a(j)} - h_{b(j)}) + (z_{b(j)} - z_{a(j)}) h_{b(j)}
    \end{split}
\end{equation*}

It is now possible to take the absolute value of both sides, and studying the different terms separately. Firstly note that we have that $|x_{b(j)}| \leq \|x_b\|_\infty \leq 1$ in $\mathcal{X}$, and that the bounds \eqref{eq:gates-bounds} hold.

Exploiting the lipschitzianity of $\sigma(\cdot)$ we can derive that
\begin{equation*}
    \begin{split}
        &|z_{a(j)} - z_{b(j)}| \leq \norm{z_a - z_b}_\infty \leq \frac{1}{4} \norm{U_z}_\infty \norm{\Delta x}_\infty
    \end{split}
\end{equation*}
In a similar way, we have that
\begin{equation*}
    \begin{split}
        &|r_{a(j)} - r_{b(j)}| \leq \norm{r_a - r_b}_\infty \leq \frac{1}{4} \norm{U_r}_\infty \norm{\Delta x}_\infty
    \end{split}
\end{equation*}
Finally, exploiting the lipschitzianity of $\tanh(\cdot)$, we have that
\begin{equation*}
    \begin{split}
        &|h_{a(j)} - h_{b(j)}| \leq \norm{h_a - h_b}_\infty \\
        & \leq \norm{U_h}_\infty \norm{r_a \circ x_a - r_b \circ x_b}_\infty \\
        & = \norm{U_h}_\infty  \norm{r_a \circ (x_a - x_b) + (r_a - r_b) \circ x_b}_\infty \\
        & \leq \norm{U_h}_\infty \left( \Bar{\sigma}_r \norm{x_a - x_b}_\infty + \norm{r_a - r_b}_\infty \right) \\
        & \leq \norm{U_h}_\infty \left( \frac{1}{4} \norm{U_r}_\infty + \Bar{\sigma}_r \right) \norm{x_a - x_b}_\infty
    \end{split}
\end{equation*}

Combining the previous inequalities we have that for all $j = 1,...,n$
\begin{equation*}
    |\Delta x_{(j)}^+| \leq \kappa_{x,j} \| \Delta x \|_\infty
\end{equation*}
where
\begin{equation*}
    \begin{split}
        \kappa_{x,j} &= z_{a(j)} + \frac{1}{4}(1 + \Bar{\phi}_h) \| U_z \|_\infty \\
        & + (1 - z_{a(j)}) \| U_h \|_\infty \left( \frac{1}{4} \| U_r \|_\infty + \Bar{\sigma}_r \right)
    \end{split}
\end{equation*}

In view of the definition of infinity norm, to prove \eqref{eq:incremental-lyap-neg-def}, i.e. $\norm{\Delta x^+}_\infty \leq \rho_s \norm{\Delta x}_\infty$, 
it is sufficient to show that $\kappa_{x,j} \leq \rho_s$ for all $j$.
In this regard we have that
\begin{equation*}
    \begin{split}
        & \frac{\kappa_{x,j} - z_{a(j)}}{1 - z_{a(j)}} \\ 
        & = \frac{1}{4} \frac{1 + \Bar{\phi}_h}{1 - z_{a(j)}} \| U_z \|_\infty + \| U_h \|_\infty \left( \frac{1}{4} \| U_r \|_\infty + \Bar{\sigma}_r \right) \\
        & \stackrel{\eqref{eq:bound-z-2}}{\leq} \nu
    \end{split}
\end{equation*}
Then for all $j = 1,...,n$
\begin{equation*}
    \begin{split}
        \kappa_{x,j} &\leq \nu (1 - z_{a(j)}) + z_{a(j)} = z_{a(j)} (1 - \nu) + \nu \\
        &\stackrel{\eqref{eq:bound-z-1}}{\leq} \bar{\sigma}_z (1 - \nu) + \nu = \nu (1 - \bar{\sigma}_z) + \bar{\sigma}_z = \rho_s
    \end{split}
\end{equation*}
This concludes the proof of \eqref{eq:incremental-lyap-neg-def}.

\textit{Condition \eqref{eq:incremental-lyap-constraints}:} Consider the $j$-th row of $L U_o (x_a - x_b)$ for $j = 1,...,q$:
\begin{equation*}
    \begin{split}
        &(L U_o)_{(j*)} (x_a - x_b) \leq \| (L U_o)_{(j*)} (x_a - x_b) \|_\infty \\
        & \leq \| (L U_o)_{(j*)} \|_\infty \| x_a - x_b \|_\infty = \| (L U_o)_{(j*)} \|_\infty V_s (x_a, x_b)
    \end{split}
\end{equation*}

\subsection{Proof of Lemma \ref{lem:Vo}}

\textit{Condition \eqref{eq:lyap-observer-neg-def}:} Let's study the time evolution of the $j$-th component of the observer estimation error, for all $j = 1,...,n$:
\begin{equation*}
    \begin{split}
        &x^+_{(j)} - \hat{x}^+_{(j)} = z_{(j)} x_{(j)} + (1 - z_{(j)}) h_{(j)} \\
        & - \hat{z}_{(j)} \hat{x}_{(j)} - (1 - \hat{z}_{(j)}) \hat{h}_{(j)} \\
        & = z_{(j)} (x_{(j)} - \hat{x}_{(j)}) + (z_{(j)} - \hat{z}_{(j)}) \hat{x}_{(j)} \\
        & + (1 - z_{(j)}) (h_{(j)} - \hat{h}_{(j)}) + (\hat{z}_{(j)} - z_{(j)}) \hat{h}_{(j)}
    \end{split}
\end{equation*}

We can now take the absolute value of both sides, and study the different terms separately. Firstly we note that $\mathcal{X}$ is a positive invariant set for $\hat{x}$, since $\hat{x}^+_{(j)}$ is computed as the convex combination of $\hat{x}_{(j)}$ and $\hat{h}_{(j)} \in (-1,1)$. Then $|\hat{x}_{(j)}| \leq \norm{\hat{x}}_\infty \leq 1$ and $|\hat{h}_{(j)}| \leq \|\hat{h}\|_\infty \leq \Bar{\phi}_h$.

Exploiting the lipschitzianity of $\sigma(\cdot)$ we can derive that
\begin{equation*}
    \begin{split}
        &|z_{(j)} - \hat{z}_{(j)}| \leq \norm{z - \hat{z}}_\infty \\
        & \leq \frac{1}{4} \norm{U_z (x - \hat{x}) - L_z (y - \hat{y})}_\infty \\
        & \leq \frac{1}{4} \norm{U_z - L_z U_o}_\infty \norm{x - \hat{x}}_\infty + \frac{1}{4} \norm{L_z}_\infty \norm{w_y}_\infty
    \end{split}
\end{equation*}
In the same way we have that
\begin{equation*}
    \begin{split}
        &|r_{(j)} - \hat{r}_{(j)}| \leq \norm{r - \hat{r}}_\infty \\
        & \leq \frac{1}{4} \norm{U_r - L_r U_o}_\infty \norm{x - \hat{x}}_\infty + \frac{1}{4} \norm{L_r}_\infty \norm{w_y}_\infty
    \end{split}
\end{equation*}
Finally, exploiting the lipschitzianity of $\tanh(\cdot)$, we have that
\begin{equation*}
    \begin{split}
        &|h_{(j)} - \hat{h}_{(j)}| \leq \norm{h - \hat{h}}_\infty \\
        & \leq \norm{U_h}_\infty \norm{r \circ x - \hat{r} \circ \hat{x}}_\infty \\
        & = \norm{U_h}_\infty  \norm{r \circ (x - \hat{x}) + (r - \hat{r}) \circ \hat{x}}_\infty \\
        & \leq \norm{U_h}_\infty \left( \Bar{\sigma}_r \norm{x - \hat{x}}_\infty + \norm{r - \hat{r}}_\infty \right) \\
        & \leq \norm{U_h}_\infty \left( \frac{1}{4} \norm{U_r - L_r U_o}_\infty + \Bar{\sigma}_r \right) \norm{x - \hat{x}}_\infty \\
        & + \frac{1}{4} \norm{U_h}_\infty \norm{L_r}_\infty \norm{w_y}_\infty
    \end{split}
\end{equation*}

Combining the previous computations we have that for all $j = 1,...,n$
\begin{equation*}
    |x^+_{(j)} - \hat{x}^+_{(j)}| \leq \kappa_{o,j} \norm{x - \hat{x}}_\infty + \kappa_{w,j} \|w_{y}\|_\infty
\end{equation*}
where 
\begin{equation*}
    \begin{split}
        \kappa_{o,j} &= z_{(j)} + \frac{1}{4} (1 + \Bar{\phi}_h) \norm{U_z - L_z U_o}_\infty \\
        & + (1 - z_{(j)}) \norm{U_h}_\infty \left( \frac{1}{4} \norm{U_r - L_r U_o}_\infty + \Bar{\sigma}_r \right) 
    \end{split}
\end{equation*}
\begin{equation*}
    \begin{split}
        \kappa_{w,j} & = \frac{1}{4} (1 + \phi_h) \| L_z \|_\infty + (1 - z_{(j)}) \frac{1}{4} \| U_h \|_\infty \| L_r \|_\infty
    \end{split}
\end{equation*}

It is now sufficient to show that $\kappa_{o,j} \leq \rho_o$ for all $j$ and $\kappa_{w,j} \leq \kappa$ for all $j$ to prove condition \eqref{eq:lyap-observer-neg-def}.
The inequality for $\kappa_{w,j}$ follows immediately considering that $1 - z_{(j)} \leq \Bar{\sigma}_z$, as already noted in \eqref{eq:bound-z-2}.
Considering $\kappa_{o,j}$, we have that
\begin{equation*}
    \begin{split}
        & \frac{\kappa_{o,j} - z_{(j)}}{1 - z_{(j)}} = \norm{U_h}_\infty \left( \frac{1}{4} \norm{U_r - L_r U_o}_\infty + \Bar{\sigma}_r \right) \\
        & + \frac{1}{4} \frac{1 + \Bar{\phi}_h}{1 - z_{(j)}} \norm{U_z - L_z U_o}_\infty \\
        & \stackrel{\eqref{eq:bound-z-2}}{\leq} \nu_o
    \end{split}
\end{equation*}
Then
\begin{equation*}
    \begin{split}
        \kappa_{o,j} &\leq \nu_o (1 - z_{(j)}) + z_{(j)} = z_{(j)} (1 - \nu_o) + \nu_o \\
        &\stackrel{\eqref{eq:bound-z-1}}{\leq} \bar{\sigma}_z (1 - \nu_o) + \nu_o = \nu_o (1 - \bar{\sigma}_z) + \bar{\sigma}_z = \rho_o
    \end{split}
\end{equation*}

\textit{Condition \eqref{eq:lyap-observer-constraint}:} The proof is similar to the proof of condition \eqref{eq:incremental-lyap-constraints} in Lemma \ref{lem:Vs}.

\textit{Condition \eqref{eq:lyap-observer-max-gain}:} Let's study $\hat{x}^+ - f(\hat{x}, u)$:
\begin{equation*}
    \begin{split}
        & \hat{x}^+ - f(\hat{x}, u) = \\
        & \hat{z}(\hat{x}, u, y) \circ \hat{x} + (1 - \hat{z}(\hat{x}, u, y)) \circ \hat{h}(\hat{x}, u, y) \\
        & - z(\hat{x}, u) \circ \hat{x} - (1 - z(\hat{x}, u)) \circ h(\hat{x}, u) \\
        & = (\hat{z}(\hat{x}, u, y) - z(\hat{x}, u)) \circ \hat{x} \\
        & + (1 - z(\hat{x}, u)) \circ (\hat{h}(\hat{x}, u, y) - h(\hat{x}, u)) \\
        & +  (z(\hat{x}, u) - \hat{z}(\hat{x}, u, y)) \circ \hat{h}(\hat{x}, u, y) \\
        & = (\hat{z}(\hat{x}, u, y) - z(\hat{x}, u)) \circ (\hat{x} - \hat{h}(\hat{x}, u, y)) \\
        & + (1 - z(\hat{x}, u)) \circ (\hat{h}(\hat{x}, u, y) - h(\hat{x}, u))
    \end{split}
\end{equation*}

Let's now take the infinity norm of both sides of this expression, and consider separately the different terms of this sum.

In view of lipschitzianity of $\sigma(\cdot)$ and of the fact that $\norm{\hat{h}(\hat{x}, u, y)}_\infty \leq \bar{\phi}_h$, we have that
\begin{equation*}
    \begin{split}
        & \norm{(\hat{z}(\hat{x}, u, y) - z(\hat{x}, u)) \circ (\hat{x} - \hat{h}(\hat{x}, u, y))}_\infty \\
        & \leq \frac{1}{4} \norm{L_z(y - \hat{y})}_\infty \norm{\hat{x} - \hat{h}(\hat{x}, u, y)}_\infty \\
        & \leq \frac{1}{4} (1 + \Bar{\phi}_h) \left( \norm{L_z U_o}_\infty \norm{x - \hat{x}}_\infty + \norm{L_z}_\infty \norm{w_y}_\infty \right)
    \end{split}
\end{equation*}

Note now that
\begin{equation*}
    \begin{split}
        &\norm{\hat{r}(\hat{x}, u, y) - r(\hat{x}, u)}_\infty \leq \frac{1}{4} \norm{L_r (y - \hat{y})}_\infty \\
        & \leq \frac{1}{4} \norm{L_r U_o}_\infty \norm{x - \hat{x}}_\infty + \frac{1}{4} \norm{L_r}_\infty \norm{w_y}_\infty
    \end{split}
\end{equation*}
Then, in view of lipschitzianity of $\tanh(\cdot)$ and of \eqref{eq:bound-z-2}
\begin{equation*}
    \begin{split}
        & \norm{(1 - z(\hat{x}, u)) \circ (\hat{h}(\hat{x}, u, y) - h(\hat{x}, u))}_\infty \\
        & \leq \norm{(1 - z(\hat{x}, u))}_\infty \norm{U_h (\hat{x} \circ \hat{r}(\hat{x}, u, y) - \hat{x} \circ r(\hat{x}, u))} \\
        & \leq \frac{1}{4} \Bar{\sigma}_z \norm{U_h}_\infty \left( \norm{L_r U_o}_\infty \norm{x - \hat{x}}_\infty + \norm{L_r}_\infty \norm{w_y}_\infty \right)
    \end{split}
\end{equation*}

Combining the previous computation we obtain that
\begin{equation*}
    \norm{\hat{x}^+ - f(\hat{x}, u)}_\infty \leq L_{max} \norm{x - \hat{x}}_\infty + \kappa \norm{w_y}_\infty
\end{equation*}
so that \eqref{eq:lyap-observer-max-gain} is proven.

\subsection{Proof of Theorem \ref{th:feasibility-stability}}
\textit{Proof of the satisfaction of constraints \eqref{eq:input-constraint} and \eqref{eq:output-constraint} for $\psi \in \mathcal{X}^{MPC}$: } 
Satisfaction of \eqref{eq:input-constraint} is obtained thanks to the constraint \eqref{eq:optimization-input-constraint} in the FHOCP. To prove satisfaction of \eqref{eq:output-constraint} first note that for $j = 1, ..., q$
\begin{equation}    \label{eq:wL-max}
    w_{L(j)} = \max_{w_y \in \mathcal{W}_y} L_{(j*)} w_y
\end{equation}
Then
\begin{equation*}
    \begin{split}
        &L y_k \stackrel{\eqref{eq:perturbed-model-output}}{=} L (U_o x_k + b_o + w_{y,k}) \stackrel{\eqref{eq:lyap-observer-constraint}}{\leq} L (U_o \hat{x}_k + b_o + w_{y,k}) \\
        & +  c_o V_o(x_k - \hat{x}_k) \stackrel{\eqref{eq:optimization-initialization} \eqref{eq:X-mpc}}{\leq} L (U_o x_{0|k} + b_o) + L w_{y,k} \\
        & + c_o \hat{e}_{o,k} \stackrel{\eqref{eq:wL-max}}{\leq} L (U_o x_{0|k} + b_o) + w_L + c_o \hat{e}_{o,k} \\
        & \stackrel{\eqref{eq:initialization_a_ji} \eqref{eq:optimization-constraint}}{\leq} h - a_0  \hat{e}_{o,k} + a_0 \hat{e}_{o,k} = h
    \end{split}
\end{equation*}

\textit{Definition and study of the candidate solution: } We define now a candidate solution for the FHOCP that will be exploited both in the proof of recursive feasibility and in the proof of stability.
Firstly, given the optimal solution of the optimization problem at time step $k$, that is $u^*_{0|k}, ..., u^*_{N-1|k}$, let's denote with $x^*_{0|k}, ..., x^*_{N-1|k}$ the associate state trajectory defined by $x^*_{i+1|k} = f(x^*_{i|k}, u^*_{i|k})$ with $x^*_{0|k} = \hat{x}_k$. Let's also define $x^*_{N+1|k} = f(x^*_{N|k}, \Bar{u})$.
Then, let's define $\Tilde{u}_{i|k+1}$ with $i = 0,...,N-1$ the candidate solution at time step $k+1$, where $\tilde{u}_{i|k+1} = u^*_{i+1|k}$ for $i = 0,...,N-2$ and $\Tilde{u}_{N-1|k+1} = \Bar{u}$.
Consider also the associate trajectory $\Tilde{x}_{0|k+1}, ..., \Tilde{x}_{N|k+1}$ defined by $\Tilde{x}_{i+1|k+1} = f(\Tilde{x}_{i|k+1}, \Tilde{u}_{i|k+1})$ with $\Tilde{x}_{0|k+1} = \hat{x}_{k+1}$. 

Note that, in view of the presence of the term $w_y$ and of the observer, in general $\hat{x}_{k+1}$ can be different from $x^*_{1|k}$, and this difference propagates along the prediction horizon so that $\Tilde{x}_{i-1|k+1} \neq x^*_{i|k}$. 
In the following we will derive some bounds on the difference between the optimal state trajectory at time step $k$ and the candidate solution at time step $k+1$.
In this regard, let's define for $i = 1, ..., N+1$
\begin{equation*}
    \varepsilon_{k+i} = \tilde{x}_{i-1|k+1} - x^*_{i|k}
\end{equation*} 

For $i = 1$ we have that $\varepsilon_{k+1} = \tilde{x}_{0|k+1} - x^*_{1|k} = \hat{x}_{k+1} - f(\hat{x}_k, u^*_{0|k})$. In view of \eqref{eq:lyap-observer-max-gain}
\begin{equation*}
    \begin{split}
        \norm{\varepsilon_{k+1}}_\infty &\stackrel{\eqref{eq:lyap-observer-max-gain}}{\leq} L_{max} \norm{x_k - \hat{x}_k}_\infty + \kappa \norm{w_{y,k}}_\infty \\
        & \stackrel{\eqref{eq:X-mpc}\eqref{eq:w-bar}}{\leq} L_{max} \hat{e}_{o,k} + \Bar{w}
    \end{split}    
\end{equation*}

Then, for $i = 2, ..., N+1$, in view of \eqref{eq:incremental-lyap-neg-def}, we have that
\begin{equation}    \label{eq:bound-epsilon}
    \begin{split}
        &\norm{\varepsilon_{k+i}}_\infty \leq \rho_s^{i-1} \norm{\varepsilon_{k+1}}_\infty \\
        & \leq \rho_s^{i-1} L_{max} \norm{x_k - \hat{x}_k}_\infty + \rho_s^{i-1} \kappa \norm{w_{y,k}}_\infty \\
        & \leq \rho_s^{i-1} ( L_{max} \hat{e}_{o,k} + \Bar{w})
    \end{split}
\end{equation}

\textit{Proof of recursive feasibility: }
We first verify that the terminal set is designed so that $x \in \mathcal{X}_f \implies L(U_o x + b_o) + w_L \leq h - a_N \hat{e}_{o} - b_N$. We first note that 
\begin{equation*}
    \begin{split}
        U_o x + b_o = U_o (x - \Bar{x}) + U_o \Bar{x} + b_o = U_o (x - \Bar{x}) + \Bar{y}
    \end{split}
\end{equation*}
Consider now the $j$-th row of $L(U_o x + b_o) + w_L$, for $j = 1,...,q$:
\begin{equation*}
    \begin{split}
        & (L(U_o x + b_o) + w_L)_{(j)} = (L(U_o (x - \Bar{x}) + \Bar{y}) + w_L)_{(j)} \\
        & = (L U_o)_{(j*)} (x - \Bar{x}) + \Bar{y}) + L_{(j*)} \Bar{y} + w_{L(j)} \\
        & \leq \norm{(L U_o)_{(j*)} (x - \Bar{x})}_\infty + L_{(j*)} \Bar{y} + w_{L(j)} \\
        & \leq \norm{(L U_o)_{(j*)}}_\infty \norm{x - \Bar{x}}_\infty + L_{(j*)} \Bar{y} + w_{L(j)} \\
        & \stackrel{\eqref{eq:terminal_constraint}}{\leq} \norm{(L U_o)_{(j*)}}_\infty \alpha + L_{(j*)} \Bar{y} + w_{L(j)} \\
        & \stackrel{\eqref{eq:def-alpha}}{\leq} h_{(j)} - \tilde{e}_o a_{N(j)} - b_{N(j)} \\
        & \leq h_{(j)} - \hat{e}_{o} a_{N(j)} - b_{N(j)}
    \end{split}
\end{equation*}
i.e. $x \in \mathcal{X}_f \implies L(U_o x + b_o) \leq h - a_N \hat{e}_{o} - b_N$.

Then, using the candidate sequence introduced in the previous part of the proof and in view of the fact that $x \in \mathcal{X}_f \implies L(U_o x + b_o) + w_L \leq h - a_N \hat{e}_{o,k} - b_N$, recursive feasibility with respect to \eqref{eq:optimization-constraint} can be proven as follows:
\begin{equation*}
    \begin{split}
        &L(U_o \tilde{x}_{i|k+1} + b_o) + w_L \\
        & \stackrel{\eqref{eq:incremental-lyap-constraints}}{\leq} L(U_o x^*_{i+1|k} + b_o) + w_L + c_s V_s(\tilde{x}_{i|k+1}, x^*_{i+1|k}) \\
        & \stackrel{\eqref{eq:optimization-constraint}}{\leq} h - a_{i+1} \hat{e}_{o,k} - b_{i+1} + c_s \norm{\varepsilon_{k+i+1}}_\infty \\
        & \stackrel{\eqref{eq:constraint_tightening_parameters} \eqref{eq:bound-epsilon}}{\leq} h - \rho_o a_i \hat{e}_{o,k} - \rho_s^i L_{max} c_s \hat{e}_{o,k} \\
        & - b_i - a_i \Bar{w} - c_s \rho_s^i \Bar{w} + c_s \rho_s^i (L_{max} \hat{e}_{o,k} + \Bar{w}) \\
        & \stackrel{\eqref{eq:eo-evolution}}{=} h - a_i \hat{e}_{o,k+1} - b_i
    \end{split}
\end{equation*}

Finally, we can prove recursive feasibility with respect to the terminal constraint \eqref{eq:optimization-terminal-constraint}, i.e. prove that $\norm{\Tilde{x}_{N|k+1} - \Bar{x}}_\infty \leq \alpha$ given that $\norm{x^*_{N|k} - \Bar{x}}_\infty \leq \alpha$:
\begin{equation*}
    \begin{split}
        &\norm{\Tilde{x}_{N|k+1} - \Bar{x}}_\infty \\
        & \leq \norm{\Tilde{x}_{N|k+1} - x^*_{N+1|k}}_\infty + \norm{x^*_{N+1|k} - \Bar{x}}_\infty \\
        & \stackrel{\eqref{eq:incremental-lyap-neg-def}}{\leq} \norm{\varepsilon_{k+N+1}}_\infty + \rho_s \norm{x^*_{N|k} - \Bar{x}}_\infty \\
        & \stackrel{\eqref{eq:bound-epsilon}}{\leq} \rho_s^N (L_{max} \hat{e}_{o,k} + \Bar{w}) + \rho_s \alpha \stackrel{\eqref{eq:condition-recursive-feasibility}}{\leq} \alpha
    \end{split}
\end{equation*}

\textit{Proof of stability: } Firstly note that the evolution of the state $\hat{e}_o$ does not depend on $w_y$, and that $\hat{e}_{o,k} \to \Bar{e}_\infty$ for $k \to \infty$.

Then, instead of studying directly the evolution of the closed loop states 
$\begin{bmatrix}
    x - \Bar{x} \\
    \hat{x} - \Bar{x}
\end{bmatrix}$,
it is possible to make a change of variables and study the equivalent system with state 
$\begin{bmatrix}
    x - \hat{x} \\
    \hat{x} - \Bar{x}
\end{bmatrix}$. The advantage is that in Theorem \ref{th:iss-observer} we have already derived the ISS property for the subsystem that describes the evolution of $x - \hat{x}$. Hence it is now sufficient to study the subsystem that describes the evolution of $\hat{x} - \Bar{x}$, considering $x - \hat{x}$ as an additional input.
To do so, we consider as candidate ISS-Lyapunov function the optimal cost of the MPC optimization, i.e.
\begin{equation*}
    \begin{split}
        V(\hat{x}_k - \Bar{x}) &= \sum_{i=0}^{N-1} \left( \|x^*_{i|k} - \bar{x} \|_Q^2 + \|u^*_{i|k} - \bar{u} \|_R^2 \right) \\
        & + s \norm{x^*_{N|k} - \Bar{x}}_{\infty}^2
    \end{split}  
\end{equation*}
We start by proving that our candidate Lyapunov function respects \eqref{eq:iss-lyap-pos-def}. We have that
\begin{equation*}
    V(\hat{x}_k - \Bar{x}) \geq \|x^*_{0|k} - \bar{x} \|_Q^2 \geq \lambda_{min}(Q) \|\hat{x}_k - \bar{x} \|_2^2
\end{equation*}
Consider now the input sequence $\Tilde{u}_{i|k} = \Bar{u}$ for $i = 0,...,N-1$, and the corresponding state trajectory $\Tilde{x}_{i|k}$. This input sequence is feasible at least for $\hat{x}_k$ close to $\Bar{x}$, i.e. for $\| \hat{x}_k - \Bar{x} \|_2 \leq \delta$ for some $\delta > 0$. Hence for $\hat{x}_k$ such that $\| \hat{x}_k - \Bar{x} \|_2 \leq \delta$ we have that
\begin{equation*}
    V(\hat{x}_k - \Bar{x}) \leq \sum_{i=0}^{N-1}  \|\Tilde{x}_{i|k} - \bar{x} \|_Q^2 + s \norm{\Tilde{x}_{N|k} - \Bar{x}}_{\infty}^2
\end{equation*}
In view of the $\delta$ISS property of the GRU model there exist $\mu \geq 0$ and $\lambda \in (0,1)$ such that
\begin{equation*}
    \|\Tilde{x}_{i|k} - \bar{x} \|_2 \leq \mu \lambda^i \|\Tilde{x}_{0|k} - \bar{x} \|_2 = \mu \lambda^i \|\hat{x}_k - \bar{x} \|_2
\end{equation*}
Therefore there exist a constant $\Tilde{b} > 0$ such that if $\| \hat{x}_k - \Bar{x} \| \leq \delta$ then
\begin{equation*}
    V(\hat{x}_k - \Bar{x}) \leq \Tilde{b} \|\hat{x}_k - \bar{x} \|_2^2
\end{equation*}
Consider now $\hat{x}_k$ such that  $\| \hat{x}_k - \Bar{x} \| > \delta$. Note that there exists $V_{max} > 0$ such that $V(\hat{x}_k - \Bar{x}) \leq V_{max}$ in $\mathcal{X}^{MPC}$. Then
\begin{equation*}
    V(\hat{x}_k - \Bar{x}) \leq \frac{V_{max}}{\delta^2} \|\hat{x}_k - \bar{x} \|_2^2
\end{equation*}
Hence we have that 
\begin{equation*}
    V(\hat{x}_k - \Bar{x}) \leq b \|\hat{x}_k - \bar{x} \|_2^2
\end{equation*}
with $b = \max \{ \Tilde{b},  \frac{V_{max}}{\delta^2} \}$, so that \eqref{eq:iss-lyap-pos-def} is verified.

We consider now the variation of the Lyapunov function between subsequent time steps, to prove that \eqref{eq:iss-lyap-neg-def} is verified:
\begin{equation*}
    \begin{split}
        & V(\hat{x}_{k+1} - \Bar{x}) - V(\hat{x}_k - \Bar{x}) \\
        & = \sum_{i=0}^{N-1} \left( \|x^*_{i|k+1} - \bar{x} \|_Q^2 + \|u^*_{i|k+1} - \bar{u} \|_R^2 \right) \\
        & + s \norm{x^*_{N|k+1} - \Bar{x}}_{\infty}^2 - \Bigg( \sum_{i=0}^{N-1} \Big( \|x^*_{i|k} - \bar{x} \|_Q^2 \\
        & + \|u^*_{i|k} - \bar{u} \|_R^2 \Big) + s \norm{x^*_{N|k} - \Bar{x}}_{\infty}^2 \Bigg)
    \end{split}
\end{equation*}

Considering the feasible solution $\Tilde{u}_{i|k+1}$, we have that the corresponding cost is larger or equal than the optimal cost at timestep $k+1$. Then
\begin{equation*}
    \begin{split}
        & V(\hat{x}_{k+1} - \Bar{x}) - V(\hat{x}_k - \Bar{x}) \\
        & \leq \sum_{i=0}^{N-1} \left( \|\Tilde{x}_{i|k+1} - \bar{x} \|_Q^2 + \|\Tilde{u}_{i|k+1} - \bar{u} \|_R^2 \right) \\
        & + s \norm{\Tilde{x}_{N|k+1} - \Bar{x}}_{\infty}^2 - \Bigg( \sum_{i=0}^{N-1} \Big( \|x^*_{i|k} - \bar{x} \|_Q^2 \\
        & + \|u^*_{i|k} - \bar{u} \|_R^2 \Big) + s \norm{x^*_{N|k} - \Bar{x}}_{\infty}^2 \Bigg) \\
        & = - \|x^*_{0|k} - \bar{x} \|_Q^2 - \|u^*_{0|k} - \bar{u} \|_R^2 \\
        & + \sum_{i=1}^{N-1} \left( \|\Tilde{x}_{i-1|k+1} - \bar{x} \|_Q^2 -  \|x^*_{i|k} - \bar{x} \|_Q^2 \right) \\
        & + \sum_{i=1}^{N-1} \left( \|\Tilde{u}_{i-1|k+1} - \bar{u} \|_R^2 -  \|u^*_{i|k} - \bar{u} \|_R^2 \right) \\
        & + \|\Tilde{x}_{N-1|k+1} - \bar{x} \|_Q^2 + \|\Tilde{u}_{N-1|k+1} - \bar{u} \|_R^2 \\
        & + s \norm{\Tilde{x}_{N|k+1} - \Bar{x}}_{\infty}^2 - s \norm{x^*_{N|k} - \Bar{x}}_{\infty}^2
    \end{split}
\end{equation*}
In this expression, all the terms related to the input cancels out in view of how the candidate input sequence was selected, except for $- \|u^*_{0|k} - \bar{u} \|_R^2$.

Considering now the state terms between time steps $k+1$ and $k+N-1$, we have that
\begin{equation*}
    \begin{split}
        & \sum_{i=1}^{N-1} \left( \|\Tilde{x}_{i-1|k+1} - \bar{x} \|_Q^2 -  \|x^*_{i|k} - \bar{x} \|_Q^2 \right) \\
        & = \sum_{i=1}^{N-1} \left( \|\varepsilon_{k+i} + x^*_{i|k} - \bar{x} \|_Q^2 -  \|x^*_{i|k} - \bar{x} \|_Q^2 \right) \\
        & = \sum_{i=1}^{N-1} \left( \| \varepsilon_{k+i} \|_Q^2 + (x^*_{i|k} - \bar{x})^\top Q  \varepsilon_{k+i} \right) \\
    \end{split}
\end{equation*}

Consider now the state terms at time steps $k+N$ and $k+N+1$, i.e.
\begin{equation*}
    \begin{split}
        & \| \tilde{x}_{N-1|k+1} - \bar{x} \|_Q^2 - s \norm{ x^*_{N|k} - \bar{x}}_{\infty}^2  + s \norm{ \tilde{x}_{N|k+1} - \bar{x}}_{\infty}^2 \\
    \end{split}
\end{equation*}
We have that
\begin{equation*}
    \begin{split}
        & \| \tilde{x}_{N-1|k+1} - \bar{x}_{k+1} \|_Q^2 = \| x^*_{N|k} - \bar{x} + \varepsilon_{k+N} \|_Q^2 \\
        & =  \| x^*_{N|k} - \bar{x} \|_Q^2 +  \| \varepsilon_{k+N} \|_Q^2 + (x^*_{N|k} - \bar{x} )^\top Q \varepsilon_{k+N}
    \end{split}
\end{equation*}
Moreover, recalling that $\Tilde{x}_{N|k+1} = f(\Tilde{x}_{N-1|k+1}, \Bar{u})$, in view of the $\delta$ISS property of the GRU network, one has
\begin{equation*}
    \begin{split}
        & s \norm{ \tilde{x}_{N|k+1} - \bar{x}}_{\infty}^2 \stackrel{\eqref{eq:incremental-lyap-neg-def}}{\leq} s \rho_s^2 \norm{ \tilde{x}_{N-1|k+1} - \bar{x}}_{\infty}^2 \\
        & = s \rho_s^2 \norm{x^*_{N|k} - \bar{x} + \varepsilon_{k+N}}_\infty^2 \\
        & \stackrel{\eqref{eq:square-infty-norm}}{\leq} s \rho_s^2 \norm{x^*_{N|k} - \bar{x}}_\infty^2 + s \rho_s^2 \norm{\varepsilon_{k+N}}_\infty^2 + \\
        & + 2 s \rho_s^2 \norm{x^*_{N|k} - \bar{x}}_\infty \norm{\varepsilon_{k+N}}_\infty
    \end{split}
\end{equation*}

Then the following inequality holds:
\begin{equation*}
    \begin{split}
        & \| \tilde{x}_{N-1|k+1} - \bar{x} \|_Q^2 - s \norm{ x^*_{N|k} - \bar{x}}_{\infty}^2 + s \norm{ \tilde{x}_{N|k+1} - \bar{x}}_{\infty}^2 \\
        & \leq \norm{x^*_{N|k} - \bar{x}}_Q^2 + s \rho_s^2 \norm{x^*_{N|k} - \bar{x}}_\infty^2 - s \norm{x^*_{N|k} - \bar{x}}_\infty^2 \\
        & + \| \varepsilon_{k+N} \|_Q^2  + (x^*_{N|k} - \bar{x} )^\top Q \varepsilon_{k+N} \\
        & + s \rho_s^2 \norm{\varepsilon_{k+N}}_\infty^2 + 2 s \rho_s^2 \norm{x^*_{N|k} - \bar{x}}_\infty \norm{\varepsilon_{k+N}}_\infty
    \end{split}
\end{equation*}
Using the fact that for a vector $v \in \mathbb{R}^n$ it holds $\norm{v}_Q^2 \leq \lambda_{max}(Q)\norm{v}_2^2$ and that $\norm{v}_\infty \leq \sqrt{n} \norm{v}_2$, we have that
\begin{equation*}
    \begin{split}
        & \norm{x^*_{N|k} - \bar{x}}_Q^2 + s \rho_s^2 \norm{x^*_{N|k} - \bar{x}}_\infty^2 - s \norm{x^*_{N|k} - \bar{x}}_\infty^2 \\
        & \leq \left( n \lambda_{max}(Q) - (1 + \rho_s^2)s \right) \norm{x^*_{N|k} - \bar{x}}_\infty^2 \stackrel{\eqref{eq:definition_s}}{\leq} 0
    \end{split}
\end{equation*}

Finally, we have that
\begin{equation*}
    \begin{split}
        & - \|x^*_{0|k} - \bar{x} \|_Q^2 - \|u^*_{0|k} - \bar{u} \|_R^2  \leq - \lambda_{min}(Q) \|\hat{x}_k - \bar{x} \|_2^2
    \end{split}
\end{equation*}

Combining all the computations we obtain that
\begin{equation*}
    V(\hat{x}_{k+1} - \Bar{x}) - V(\hat{x}_k - \Bar{x}) \leq - c \|\hat{x}_k - \bar{x} \|_2^2 + \gamma(k)
\end{equation*}
where $c =  \lambda_{min}(Q)$ and 
\begin{equation*}
    \begin{split}
         \gamma(k) &= \sum_{i=1}^{N-1} \left( \| \varepsilon_{k+i} \|_Q^2 + (x^*_{i|k} - \bar{x})^\top Q  \varepsilon_{k+i} \right) \\
         & + \| \varepsilon_{k+N} \|_Q^2  + (x^*_{N|k} - \bar{x} )^\top Q \varepsilon_{k+N} \\
         & + s \rho_s^2 \norm{\varepsilon_{k+N}}_\infty^2 + 2 s \rho_s^2 \norm{x^*_{N|k} - \bar{x}}_\infty \norm{\varepsilon_{k+N}}_\infty
    \end{split}
\end{equation*}

In view of the bounds for $\varepsilon_{k+i}$ derived in \eqref{eq:bound-epsilon} and on the boundedness of the states of the GRU, there exist functions $\gamma_1, \gamma_2 \in \mathcal{K}$ such that
\begin{equation*}
    \begin{split}
        &V(\hat{x}_{k+1} - \Bar{x}) - V(\hat{x}_k - \Bar{x}) \leq - c \|\hat{x}_k - \bar{x} \|_2^2 \\
        & + \gamma_1(\|x_k - \hat{x}_k\|) + \gamma_2(\|w_{y,k}\|)
    \end{split}
\end{equation*}
i.e. $V(\hat{x}_k - \Bar{x})$ is an ISS-Lyapunov function for the subsystem describing the evolution  of $\hat{x} - \Bar{x}$, considering as inputs $x - \hat{x}$ and $w_y$. 

To sum up, we have shown that the closed loop system can be described by two connected subsystems. The subsystem that describes the evolution of $x - \hat{x}$ is ISS with respect to the input $w_y$, and is independent by the subsystem describing the evolution of $\hat{x} - \Bar{x}$. The subsystem that describes the evolution of $\hat{x} - \Bar{x}$ is ISS considering as inputs $x - \hat{x}$ and $w_y$. Then, in view of \cite[Theorem 2]{Jiang2001iss}, the overall system composed by the connection of the two subsystems is ISS with respect to the input $w_y$.

\end{document}